\documentclass[amssymb,amsfonts,aps,prl,twocolumn,longbibliography]{revtex4}

\usepackage{graphicx}
\usepackage{dcolumn}
\usepackage{bm}
\usepackage[normalem]{ulem}
\usepackage{xcolor}
\usepackage{comment}
\usepackage{amsmath}
\usepackage{siunitx}
\usepackage{dsfont}
\usepackage{mathrsfs}  

\usepackage{float}
\usepackage{placeins}
\usepackage[shortlabels]{enumitem}

\DeclareSIUnit\angstrom{\text{Å}}

\begin{document}

\title{Theory of atomic-scale direct thermometry using  ESR-STM }

\author{Y. del Castillo$^{1,2}$, J. Fern\'{a}ndez-Rossier$^1$\footnote{On permanent leave from Departamento de F\'{i}sica Aplicada, Universidad de Alicante, 03690 San Vicente del Raspeig, Spain}$^,$\footnote{joaquin.fernandez-rossier@inl.int} }

\affiliation{$^1$International Iberian Nanotechnology Laboratory (INL), Av. Mestre Jos\'{e} Veiga, 4715-330 Braga, Portugal }
\affiliation{$^2$Centro de F\'{i}sica das Universidades do Minho e do Porto, Universidade do Minho, Campus de Gualtar, 4710-057 Braga, Portugal }
\date{\today}

%-----------------------------------------------------------------------
\begin{abstract} 
Knowledge of the occupation ratio and the energy splitting of a two-level system yields a direct readout of its temperature. 
Based on this principle,  the determination of the temperature of an individual two-level magnetic atom was demonstrated  using 
Electron Spin Resonance (ESR) via Scanning Tunneling Microscopy (ESR-STM). The temperature determination proceeds in two steps. First, energy splitting is determined using ESR-STM. Second, the equilibrium occupation of the two-level atom is determined in a resonance experiment of a second nearby atom, that has now two different resonant peaks, associated to the two states of the magnetic two-level atom. The ratio of the heights of its resonance peaks yields the occupation ratio. Here we present theory work to address three key aspects: first, we find how shot-noise and back-action limit the precision of this thermometry method; second, we study how the geometry of the nearby spins can be used to enhance signal-to-noise ratio. We predict  ESR-STM thermometry can achieve a resolution of 10 mK using temperatures in the T= 1K range. Third, we show how ESR-STM thermometry can be used to detect thermal gradients as small as 5 mK/nm.

\end{abstract}

\maketitle

%------------------------------------------------------------------------
%\section{Introduction}

Temperature measurements play a central role in many fields of science and technology \cite{childs00}. A large variety of techniques are used to measure temperature, that depend on different physical phenomena and are better suited for different ranges.  These include resistance thermometry, that exploits the predictable change of electrical resistance with temperature of materials, such as platinum\cite{childs00}, magnetic thermometers that use the Curie-Weiss law\cite{dellby71}, noise thermometry that relies on the well-established scaling of Johnson-Nyquist noise with temperature\cite{white96,qu19},  nuclear quadrupole resonance thermometry \cite{utton67}, etc. 
A subclass of these methods is specifically designed to measure temperature with nanoscale probes\cite{brites12}, aiming to map temperature with very high spatial resolution, with important applications for efficient heat management in nanoelectronics\cite{zhang20} and in biomedical applications\cite{aslam2023}. Important nanoscale thermometers include NV center magnetometry, which exploits the shift of resonance frequency due to thermal expansion \cite{fukami19}, single-molecule chromophores, exploiting the temperature dependence of the photoluminescence linewidth\cite{esteso2023}, superconducting scanning tunnel microscope (STM) junctions\cite{esat23} and scanning thermal magnetometry\cite{zhang20}.

All of these methods infer temperature by exploiting the well-known relation between a given physical property (resistance, resonance frequency, magnetization, noise) and temperature. Therefore, it can be said that they are {\em indirect}. In contrast, a direct determination of temperature would rely on its definition in terms of the average energy of particles in a system. In the context of a quantum system with two energy levels, $G$ and $X$, temperature is defined in terms of the ratio of their occupations, following the Gibbs-Boltzmann equation: 
\begin{equation}
r=\frac{p_X}{p_G}=e^{-\beta\varepsilon},
\end{equation}
where $\varepsilon\equiv\varepsilon_X-\varepsilon_G$. Therefore,  measurement of both  $\varepsilon$ and $r$  would directly yield:
\begin{equation}
k_B T = -\frac{\varepsilon}{\ln{r}},
\label{eq:temprelation}
\end{equation}
thereby providing a {\em direct} measurement of temperature\footnote{Or more precisely, temperature times the Boltzmann constant $k_B$}, based on its definition.  Building on the experimental work of Choi {\em et al.}\cite{choi17}, here we provide a theory for the resolution and the range of the method, and we also go beyond the original work by proposing an extension to measure thermal gradients. 

The work of Choi {\em et al.}\cite{choi17} relies on ESR-STM (Electron Spin Resonance with Scanning Tunneling Microscopy) \cite{baumann15,yang17,willke18,natterer19,seifert20,van21,farinacci22,kovarik22,zhang22,kot23,wang23} in a lateral sensing scheme to determine the temperature of individual magnetic atoms or molecules placed on a surface, whose location is determined with sub-${\rm \AA}$ngström resolution. Therefore, this method vastly outperforms the spatial resolution of any other scanning thermal microscopy techniques \cite{zhang20}.

The lateral sensing scheme entails an ESR-STM active surface spin, either an atom or a molecule, that we shall call the sensor spin  (black atom under the STM tip in Figure 1a), placed nearby a second spin or group of spins, that we shall call the probe spin(s) (red atom in Figure 1a). The resonance spectrum of the sensor spin features several resonance peaks (see Figure  1b), instead of only one, on account of the dipolar coupling to the probe spins. The relative intensity of the different sensor resonance peaks reflects the relative probability for a given probe spin state to be occupied. Experiments\cite{choi17} have shown that the occupation of the probe states follows a thermal distribution, in contrast with that of the sensor atom, governed by the ESR-STM drive. This observation provides strong evidence of a very reduced back action of the sensor spin on the occupation of the probe-spin states, essential in the following discussion.
\begin{figure}
    \centering
    \includegraphics[width=1\linewidth]{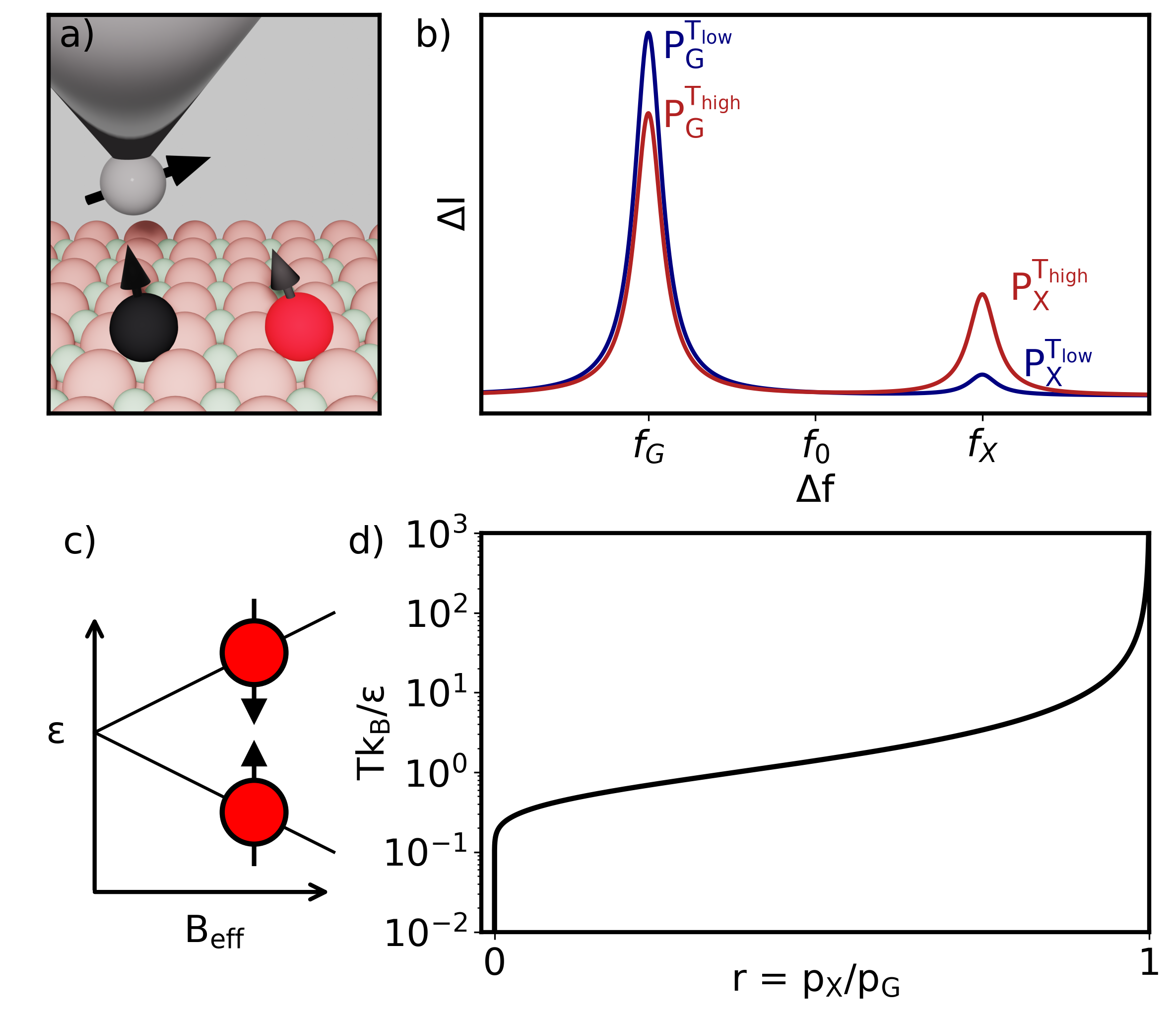}
    \caption{a) Scheme of the proposed atom arrangement to obtain the temperature. An STM tip over the sensor atom under the influence of the stray field generated by a nearby atom. b) ESR spectrum showing two resonance peaks corresponding to each state of the nearby atom for two different temperatures and how the ratio between the heights of the peaks changes. c) Energy splitting, $\varepsilon$, between the ground and excited states of the probe atom. d) Temperature dependence of the ratio between the up and down state peaks. }
    \label{fig:Fig1 scheme}
\end{figure}

We first discuss the simplest case where an ESR-STM active surface spin is coupled to an individual $S=1/2$ surface spin (See Figure \ref{fig:Fig1 scheme}a). We assume that the external magnetic field is dramatically larger than the dipolar and exchange interactions between the ESR-STM active sensor spin and the probe spin, whose temperature is being determined, so that the spin-flip terms in the Hamiltonian are negligible:
\begin{equation}
    H= g_{s}\mu_B B S_z(s)  +  g_{p}\mu_B B S_z(p) +  j S_z(s)S_z(p)
\end{equation}
where $s$ and $p$ stand for sensor and probe spin respectively. Then, we can write the sensor Hamiltonian as: 
\begin{equation}
    H_{\rm sensor}= g_{s}\mu_B B_{eff} S_z({s})  
\end{equation}

where 
\begin{equation}
 B_{eff}= B+ \frac{\mu_0 \mu_B g_p S_z(p)}{4 \pi d_{s-p}^3}.
\end{equation}
%\begin{equation}
% B_{eff}= B+  %\frac{j S_z({s} )}{g_{s}\mu_B}
%\end{equation}
%\bluemark{where $j=F(g_s,g_p, d)$}.
% g mu_B B_eff = \mu_0/4 pi d^3 g mu_B s 
%B_eff= \mu_0/(4 Pi g mu_B d^3)
Since the effective field can take 2 values, depending on whether the probe spin is in the state up or down, the ESR-STM spectrum for the sensor spin has two resonance peaks (see Figure \ref{fig:Fig1 scheme}c and b respectively). 
%We assume 
Experiments\cite{choi17} show that the ESR-STM spectra of the sensor spin relate to the occupation of the states of the probe spins, according to the equation
\begin{equation}
I(f)= p_{G} L(f-f_{G})+ p_{X} L(f-f_{X})
\end{equation}
where $p_{G,X}=\frac{1}{Z}e^{-\beta \varepsilon_{G,X}}$ are the thermal occupations of each state and $L(f-f_{G,X})$ is a Lorentzian type resonance curve centered around the frequency $f_{G,X}$. Assuming that the external field and the stray field of the probe atom only have perpendicular components, $f_{G,X}$ is given by
\begin{equation}
f_{G,X}=\frac{\mu_B g_{\rm s}}{h} B_{eff}(G,X),
\label{eq:wsen}
\end{equation}
where  $h=2\pi \hbar$, and $g_{\rm s}$ is the gyromagnetic factor of the sensor.

Equation (\ref{eq:temprelation}) relates the temperature to two quantities, the energy splitting of the probed spin, $\varepsilon$, and the ratio of the heights of the peaks in the ESR-STM spectrum of the sensor spin, $r$. These quantities have to be determined independently. The ratio $r$ is obtained from the ESR-STM spectrum of the sensor atom. The most optimal determination of $\varepsilon$ would be carried out using  ESR-STM on the probe spin, as we discuss below, given the very good spectral resolution of this method, compared to inelastic electron tunnel spectroscopy (IETS). The precision of the determination of temperature using this method is thus limited by the precision to determine $\varepsilon$ and $r$.

\begin{equation}
    \Delta T= \left| \frac{\partial T}{\partial \varepsilon} \right|\Delta \varepsilon
    +\left| \frac{\partial T}{\partial r} \right| \Delta r
    \label{eq:breakdown}
\end{equation}

After derivation and rearrangement of the resulting equation, we arrive at the following expression (See Supplemental Material \nocite{del23,delgado17}\cite{supp}):
\begin{equation}
       \frac{\Delta T}{T} =  \frac{\Delta \varepsilon}{\varepsilon} +\frac{1}{\sqrt{N}} \frac{1+r}{r   |\ln{r}|}\equiv
        \frac{\Delta \varepsilon}{\varepsilon} +\frac{1}{\sqrt{N}} {\cal F}(r)
\label{eq:resolution}
\end{equation}
where $N=\frac{I \Delta t}{e}$ is the number of tunneling electrons during the  measurement time $\Delta t$.
A more transparent expression is obtained if we use $\beta\epsilon$ instead of $r$:
\begin{equation}
       \frac{\Delta T}{ T} = \frac{\Delta \varepsilon}{\varepsilon} +
      \frac{1}{\sqrt{N}}\frac{1+e^{\beta\varepsilon }}{\beta \varepsilon}\equiv
      \frac{\Delta \varepsilon}{\varepsilon} +
      \frac{1}{\sqrt{N}}{\cal F}(\beta \epsilon)
\label{eq:resolution1}
\end{equation}
We now assume that $\frac{\Delta \varepsilon}{\varepsilon}$ is determined with ESR-STM. 
There are two sources of error for the determination of $\varepsilon$, that we label as intrinsic and back-action. The intrinsic sources of error relate to the experimental measurement of $\varepsilon$  obtained in the absence of back-action by performing a spin resonance measurement on the isolated probe atom (if this one is ESR active). Using the result from \cite{del24}, the shot-noise limited expression for the absolute error in $\varepsilon$ is:
\begin{equation}
%\Delta \varepsilon= \frac{4}{3} \frac{\hbar \delta f}{ \sqrt N_{\varepsilon}}
\frac{\Delta \varepsilon |_{\rm intrinsic}}{\varepsilon}= \frac{4}{3} \frac{1}{ \sqrt N_{\varepsilon}}
 \frac{ \delta f}{f}
 \label{eq:intrinsic}
\end{equation}
where $\delta f$ is the width of the ESR-STM spectrum resonance peak and $N_{\varepsilon}$ is the number of tunneling electron during the measurement of $\varepsilon$, that may or may not be the same than $N$ in eq. \ref{eq:resolution}.  The intrinsic contribution to the error of  $\varepsilon$ has a prefactor controlled by the quality factor of the resonance,  $ \frac{ \delta f}{f}$  for which values smaller than  $10^{-4}$ have been reported \cite{baumann15}. Therefore, we anticipate that the contribution of eq. (\ref{eq:intrinsic}) is going to be negligible.

We now discuss the role of the back-action of the sensor spin on the occupation of the probe spins.  Even if flip-flop interactions are blocked, on account of the small ratio between their magnitude and the energy difference of the states $G_sX_p$ and $X_sG_p$, the Ising interaction $j S_z(s) S_z(p)$ is still active. Therefore, the effective field of the probe spin is given by the sum  of the external field and the stray field created by the sensor spin, that depends on the average magnetization of the sensor spin,$\langle S_z(s)\rangle$, ie, $B_{\rm eff, p}= B + B_{\rm stray}(\langle S_z(s)\rangle)$
At resonance, and in the limit $T_1 T_2 \Omega^2>>1$,  we have  $\langle S_z(s)\rangle \rightarrow 0$, where $T_1, T_2, \Omega$ are the spin relaxation time, spin decoherence time and Rabi driving force of the sensor spin (See Supplemental material \cite{supp}). However, given that the resonance value $\delta=0$ can only be achieved a fraction of the instances, on account of the fact that $\delta$ depends on the state in which the probe spin is, the effective field of the probe spin is {\em definitely different} from the external field by an amount {\em smaller}
than $\frac{j S(s)}{g_p\mu_B}$.  We can therefore provide the following bound to $\varepsilon$
\begin{equation}
    \Delta \varepsilon|_{\rm back \, action}\leq jS(s) \propto B_{\rm stray}
\end{equation}
taking the most pessimistic assumption that the average spin of the sensor is maximal. 
The relative error is given by the ratio of the external field and the stray field:
\begin{equation}
\frac{    \Delta \varepsilon|_{\rm back \, action}}{\varepsilon} \leq \frac{B_{\rm stray}}{B} \simeq 10^{-3}
\label{eq:BA}
\end{equation}
Therefore, as anticipated, the back-action error is clearly larger than the intrinsic shot-noise limit (eq. \ref{eq:intrinsic}) when it comes to determining $\Delta \varepsilon$. 

\begin{figure}[h!]
    \centering
    \includegraphics[width=1\linewidth]{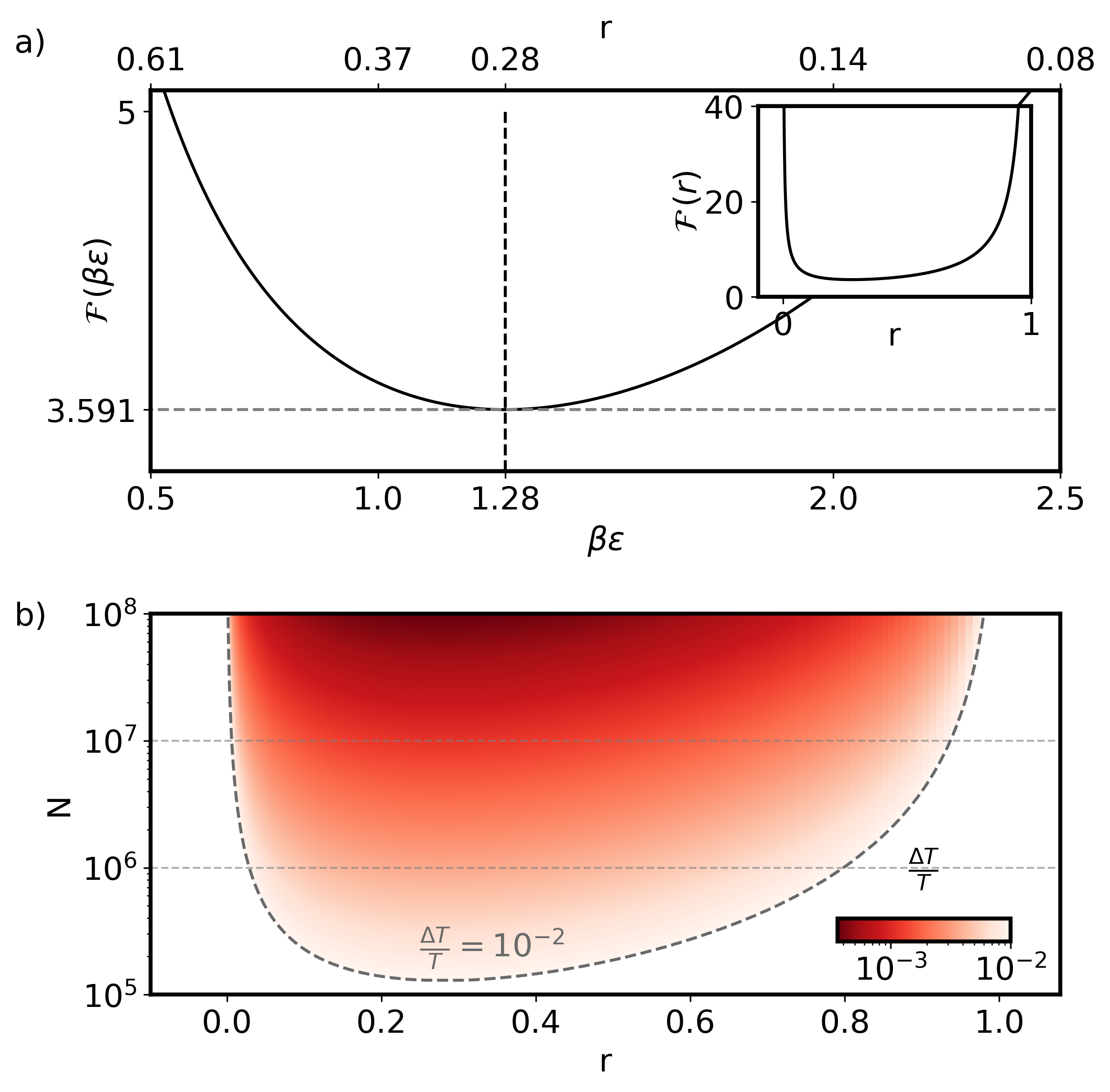}
    \caption{a) Numerical analysis of the minimal value of the second term in eq. \ref{eq:resolution} as a function of $r$ (top axis in red) and analogously for eq. \ref{eq:resolution1} as a function of $\beta \varepsilon$. Inset shows the same result for the complete range of $r$, from 0 to 1. b) Range of r where the relative error in temperature is at least $10^{-2}$ or lower as a function of N.}
    \label{fig:Fig2}
\end{figure}

We now assess the magnitude of the shot-noise limit of $\Delta r$  (second term in eq. \ref{eq:resolution}) and
compare it to the back-action error (eq. \ref{eq:BA}). In  Figure \ref{fig:Fig2}a, we show the function ${\cal F}$ that controls $\Delta r$  has a minimum ${\cal F}=3.59$ for  $r\approx 0.28$ ($\beta \varepsilon \approx 1.28$) then grows very rapidly away from that minimum (see inset), diverging both for $r\rightarrow 0$ and $r\rightarrow 1$. Therefore, the operational range of the ESR-STM thermometry is governed by the range of $r$ in the neighborhood of  $r\simeq 0.28$ where the function ${\cal F}$ remains small.

We now estimate the minimal and maximal temperatures that could be measured, if we impose an upper bound to the second contribution to error in eq. (\ref{eq:breakdown}),  
$\frac{\Delta T}{T}|_{\Delta r}= \frac{1}{\sqrt{N}}{\cal F}(r) \leq 10^{-2}$. For the sake of this estimate, we assume that the measurement time is $\Delta t\approx 2 s$, and an excess current of $\Delta I=100$ fA \cite{baumann15,willke18-2,natterer17,willke19}, this leads to $N\approx 1.25 \cdot 10^6$. We find the minimal and maximal values of  $\beta\varepsilon$ are 0.2 and 3.7 respectively.
We note that the operational range of the ESR-STM thermometer range depends on the applied magnetic field.
Assuming an applied field of 1T, standard in the ESR-STM \cite{baumann15,choi17,natterer17,willke18,willke19},  and $g=2$, we have $\varepsilon=115.7 \mu eV$, so that the measurable temperature range with a relative error of $\frac{\Delta T}{T} \leq 10^{-2}$ is between 0.36 and 6.77 K. In contrast, if ESR-STM is carried out at 0.1T\cite{steinbrecher21}, our upper bound for $\frac{\Delta T}{T}|_{\Delta r}$ is satisfied for temperatures between 36 and 677 mK. However, it must be noted that this would in turn increase the back-action error (eq. (\ref{eq:BA})).

In summary, three factors limit the resolution of the ESR-STM. According to our estimates, the most important one is the back-action error due to the stray field of the sensing spin on the probe spin, that brings stochastic changes to the energy splitting $\varepsilon$ that enters in eq. (\ref{eq:temprelation}).  
The shot-noise limits the peak-height ratio determinations. This error can be made smaller than the back-action error by increasing the measurement time, via the $1/\sqrt{N}$ prefactor. In addition, this error determines the operational range of the ESR-STM thermometer, on account of the temperature dependence of ${\cal F}(\beta\varepsilon)$ function (See Figure \ref{fig:Fig2}b). Finally, the resolution limit to determine $\varepsilon$ is negligible, compared to the other two.

We now discuss additional limitations of the ESR-STM thermometry. First, there is an upper bound to the  Zeeman splitting $\varepsilon$ on the isolated probe atom that can be probed via ESR-STM, given that it is limited by the maximum driving frequency. State of the art in an ESR-STM experiment is approximately 100 GHz \cite{drost22}. For a spin-1/2 sensor with one $\mu_B$, this corresponds to a maximum external magnetic field of 3.6 T and an energy splitting of 413.6 $\mu eV$.
Two factors limit the determination of the ratio of peak heights associated with the two magnetic states of the probe atom. First, the magnitude of the peak splitting, determined by the magnitude of the stray field at the sensor spin, has to be larger than the peak linewidth, so that two peaks can be resolved.
Consequently, this sets an upper limit to the separation of the sensor and the probe atoms. Shot noise limits the smallest variation of peak heights that can be resolved\cite{del24}. Whereas shot noise can be theoretically reduced by increasing the measurement time, thermal drift of the STM position also sets an upper limit for this quantity. Displacements on the order of picometers per hour have been reported \cite{choi17}. This would lead to changes in the stray field and therefore a shift of the resonance peak.
Finally, another degree of freedom to consider is the contribution of the stray field generated by the tip, where this one can have a very large magnetic moment. Although the stray field decays with the cube of the distance and can be thoroughly studied, it could vary in magnitude and direction depending on different external magnetic fields.

We now discuss how having ${\cal N}$ probe spins symmetrically placed around a sensor spin can be used to reduce shot noise in temperature readout. This concept was implemented by Choi and coworkers \cite{choi19}. First, we discuss the idealized case where the ${\cal N}$ probe spins do not interact with each other, have the same temperature, and their stray fields at the sensor spin are identical.

The ratio between the two lower energy peaks, the ground state and the one corresponding to flipping one spin (n=0 and n=1, respectively), is given by the following expression (see supplementary material \cite{supp}):
\begin{equation}
r=\frac{P_1}{P_0}= {\cal N}e^{-\beta \varepsilon}
\end{equation}
From here, we derive the temperature equation:
\begin{equation}
k_B T = -\frac{\varepsilon}{\ln \left( \frac{r}{{\cal N}} \right)}
\end{equation}

Analogously to the derivation of equation \ref{eq:resolution}, for ${\cal N}$ identical probe spins we find:
\begin{equation}
{\cal F}_{{\cal N}}(r) =\frac{r+1}{r |\ln{(r/{\cal N})}|}  \text{  and  } {\cal F}_{{\cal N}}(\beta \varepsilon) =\frac{{\cal N} + e^{\beta \varepsilon}}{ {\cal N} \beta \varepsilon} 
\label{eq:N-Resolution}
\end{equation}
\begin{figure}[h!]
    \centering
    \includegraphics[width=1\linewidth]{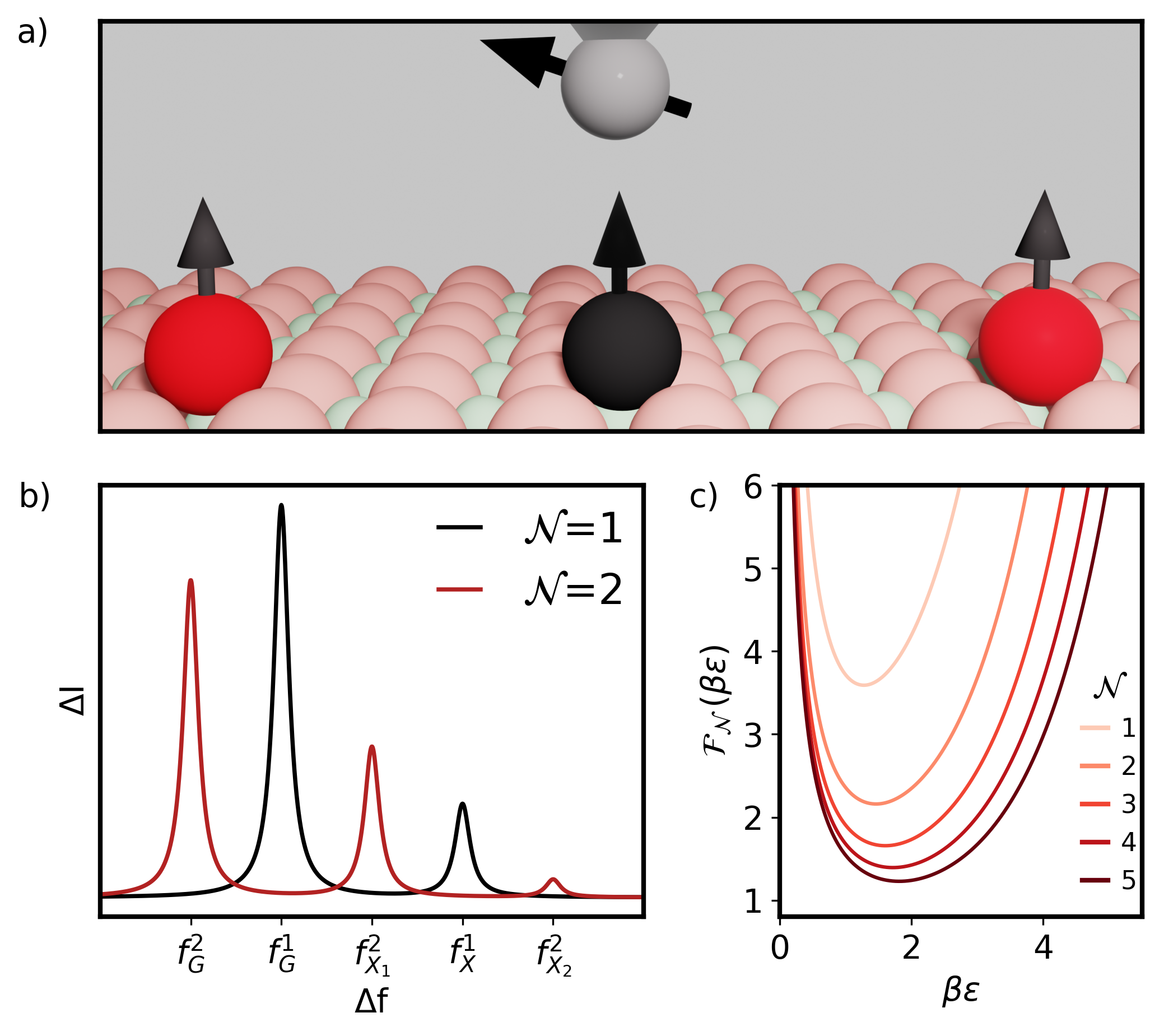}
    \caption{a) Proposed arrangement using ${\cal N}=2$ identical probe spins to improve the range and precision of measurements. b) Comparison between the ESR spectrum using one and two probe spins. c) Minima of ${\cal F}_{\cal N}$ as a function of ${\cal N}$.}
    \label{fig:Fig3}
\end{figure}
For the simplest case of an engineered structure with ${\cal N}=2$, a sensor is placed equidistant between two identical atoms. We can label the four states of the probe atoms as $GG$, $GX$, $XG$ and $XX$, where $G$ and $X$ stand for ground and excited state, respectively. The complete ESR spectrum consists of three peaks: two lateral peaks, for states $GG$ and $XX$ and a central peak at the resonance frequency of the isolated sensor atom, that corresponds to the states $GX$ and $XG$, whose stray fields cancel each other. The height of the central peak is proportional to the joint probability of the two possible states, resulting in almost twice the change in height for the same temperature variation compared to a single-atom readout.

This improvement in the variation of the peak with temperature translates to a reduced contribution to the $\Delta r$ noise (second term in eq. (\ref{eq:breakdown})).
As shown in Figure \ref{fig:Fig3}b, compared to a single atom, for ${\cal N} = 2$, the measurable temperature range within the divergence points of ${\cal F}_{{\cal N}}$ is approximately 1.5 times larger, and the minimum relative error is about two-thirds.

In Figure \ref{fig:Fig3}b and c, we show how the $\beta \varepsilon$ range allowed for measurement and the minimum presented in eq. \ref{eq:resolution}, respectively, improve with ${\cal N}$. We observe that the improvement saturates and, for most practical scenarios, $2 \leq {\cal N} \leq 5$ is sufficient.

\begin{figure}[h!]
\includegraphics[width=1\linewidth]{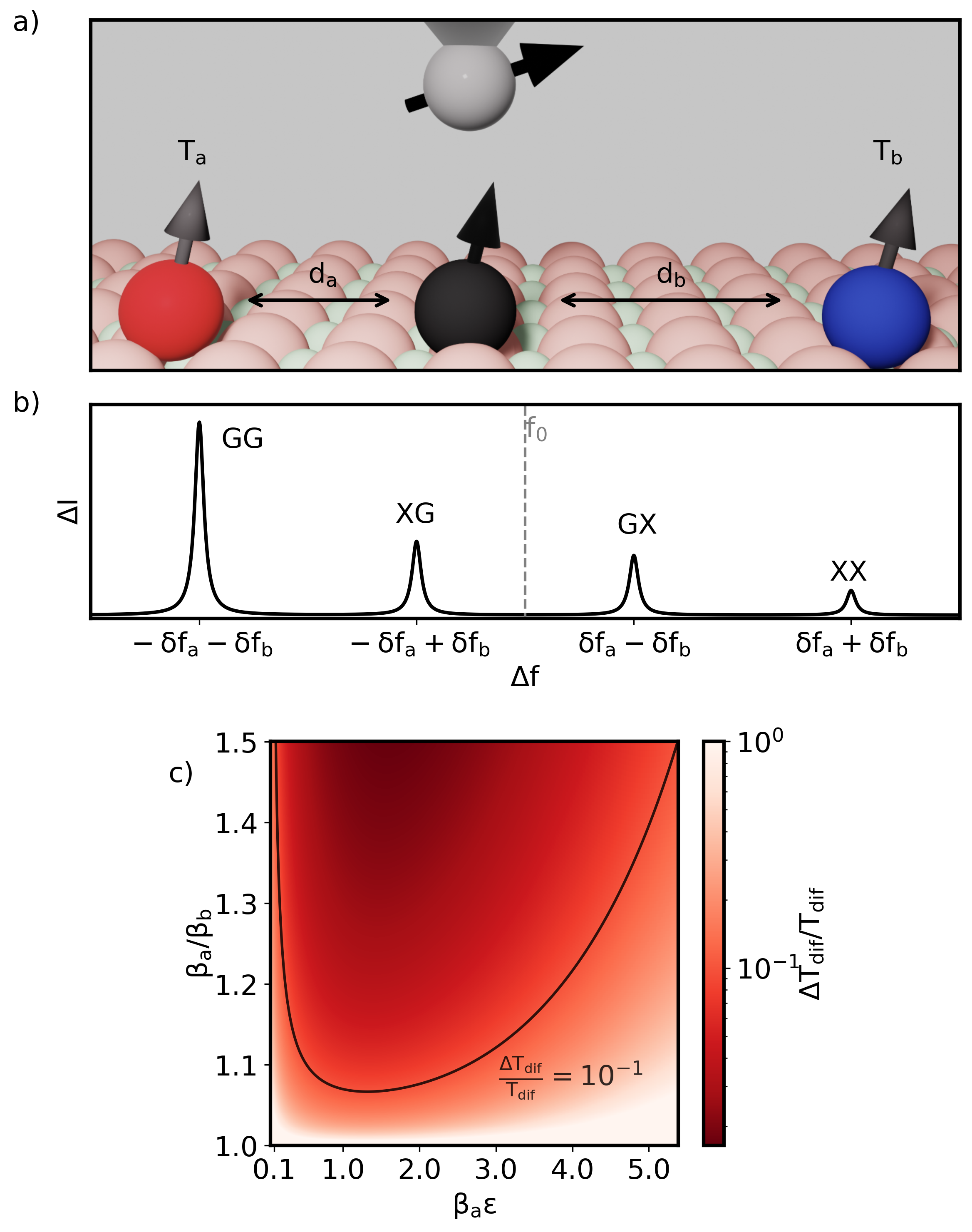}
    \caption{a) Scheme proposed to measure thermal gradient. A sensor atom is placed non-equidistantly between two atoms with different temperatures. b) ESR spectrum resulting from the atom arrangement in panel a. c) Relative error of the thermal difference as a function of both, the temperature in one spin and the difference between both probe spins for $N=1.25\cdot 10 ^6$.}
\label{fig:gradient}
\end{figure}

We now propose a method to probe thermal gradients in two atoms, extending the ESR-thermometry to the case where temperatures are not homogeneous. We consider an ESR active sensor spin placed between two identical two-level magnetic atoms, that we label as $a$ and $b$, whose temperature difference $T_a-T_b$ we want to determine (see Figure \ref{fig:gradient}a). The thermal gradient is given by 
\begin{equation}
\frac{dT}{dx}
%\Delta T
\simeq \frac{T_a-T_b}{d_a-d_b}.
\end{equation}
where $d_a$ and $d_b$ are the positions of $a$ and $b$ atoms along the $x$ axis in the plane of the surface.
The atoms $a$ and $b$ are separated by a distance in the range of 2-4 nanometers. This distance is limited by the range of the stray field created by a point magnetic moment and the spectral sensitivity of the ESR active spin. The values of $d_a$ and $d_b$ can be determined using STM with subatomic resolution. Therefore, the non-trivial part of the thermal gradient measurement is the determination of the thermal difference
$ T_{\rm dif}\equiv T_a-T_b$.
To do so, we propose to place the sensor atom in the line that joins atoms $a$ and $b$, but closer to one of the probe atoms. This differs from the geometry considered in Figure \ref{fig:Fig3}a, where the sensor atom was equidistant from the outer atoms. As a result, instead of a single central peak that corresponds to the $XG$ and $GX$ configurations, there are two peaks, as the stray fields no longer compensate each other (see Figure \ref{fig:gradient}b) and a total of four peaks. Since we assume again that the magnetic moments of the spins are along the $z-$axis, perpendicular to the surface, the external and stray fields add up, and the frequencies of the four resonant peaks are given by:
\begin{equation}
    f_{\sigma_a \sigma_b}= f_0 + \sigma_a \delta f_a + \sigma_b \delta f_b
\end{equation}

The determination of the thermal difference is very similar to the determination of temperature. We first realize that we can easily relate the temperature variation to the exponential factors that appear in Boltzmann distributions:
\begin{equation}
    T_{dif} = -\frac{\varepsilon}{k_B} \bigg ( \frac{1}{ln R_a } - \frac{1}{ln R_b} \bigg )
    \label{eq:Tgrad}
\end{equation}
where $R_a=e^{-\beta_a \varepsilon}$ and $R_b=e^{-\beta_b \varepsilon}$. We next relate $R_a$ and $R_b$ to the ratios of the heights of the peaks in the ESR-STM spectrum of the sensor atom.
To derive the heights of the peaks of the ESR-STM spectra, we assume that Zeeman energy is much larger than dipolar (or exchange) coupling between atoms $a$ and $b$. As a result, the energy spectra are additive, $E_{GG}=-\frac{\varepsilon_a+\varepsilon_b}{2}$,
$E_{GX}=-\frac{\varepsilon_a-\varepsilon_b}{2}$
$E_{XG}=-\frac{\varepsilon_b-\varepsilon_a}{2}$
$E_{XX}=+\frac{\varepsilon_a+\varepsilon_b}{2}$. Since the atoms are identical, we take $\varepsilon_a=\varepsilon_b=\varepsilon$.  
We can now write the Boltzmann factors for the states $GG, GX, XG$  as products 
of probabilities of atoms $a$ and $b$: 
\begin{eqnarray}
P_{GG}=  \frac{e^{\beta_a \frac{\varepsilon}{2}} e^{\beta_b \frac{\varepsilon}{2}}}{Z}
 \nonumber\\
P_{XG}= \frac{e^{-\beta_a \frac{\varepsilon}{2}} e^{\beta_b \frac{\varepsilon}{2}}}{Z}
 \nonumber\\
P_{GX}= \frac{e^{\beta_a \frac{\varepsilon}{2}} e^{-\beta_b \frac{\varepsilon}{2}}}{Z}
\end{eqnarray}
where $Z$ is the partition function.  From here we obtain right away:
\begin{eqnarray}
    R_a=\frac{P_{XG}}{P_{GG}}; \;\; 
    R_b=\frac{P_{GX}}{P_{GG}} \;\; 
\label{eq:ratio-gradient}
\end{eqnarray}

Therefore, the determination of the temperature difference is then carried out from the readout of the ratios of the two central peaks in the ESR spectra with the ground state ($GG$) peak.  The analysis of the resolution of the proposed
ESR-STM thermal gradient measurement
 runs parallel to the one of the ESR-STM thermometer, discussed above. Analogously to eq. \ref{eq:resolution} and \ref{eq:resolution1} we find the following expression for the temperature difference resolution (See Supplemental Material \cite{supp}):
\begin{equation}
    \frac{\Delta T_{dif}}{T_{dif}} = \frac{\Delta \varepsilon}{\varepsilon}   + \frac{1}{\sqrt{N}}{\cal F}_{T_{dif}} (R_a, R_b)
    \label{eq:errgrad}
\end{equation}
where 

\begin{equation}
\begin{split}
{\cal F}_{T_{dif}} (R_a, R_b)=& \frac{ln (R_a)ln (R_b) }{  ln \left (R_b/ R_a\right )} \bigg ( \frac{1 + R_a}{R_a ln^2 (R_a) } + \\
&  \frac{1 + R_b}{R_b ln^2 (R_b) } \bigg)
\end{split}
\end{equation}
and
$N$ is the number of tunneling electrons recorded during the measurement time.  
The first contribution to the error is dominated by back-action error (see eq. (\ref{eq:BA})).
The second contribution, associated to the shot-noise limit in the determination of peak-height ratios, is shown  in 
figure \ref{fig:gradient}c, 
as a function of the temperature of spin $a$ ($\beta_a \varepsilon$) and the temperature ratio of atoms $b$ and $a$ ($\beta_a/\beta_b$)
We assume 
$N = 1.25 \cdot 10^6$, that corresponds to a measurement time of approximately 2 seconds, with a current of 100 fA. The solid line in the figure corresponds to a shot-noise relative error of 10 percent.  From our numerics,  the shot-noise error is never smaller than 1 percent (for the assumed value of $N$), and therefore, larger than the first term in eq. (\ref{eq:errgrad}). We 
identify an optimal temperature range, centered around $\beta_a \varepsilon \approx 1.28$, where the relative error is minimized. The relative error also increases as the temperature difference between the atoms decreases. 

To give an example of the operational temperature range of the STM-ESR temperature gradient measurement, we assume that  
atoms $a$ and $b$ have $g=2$, $S=1/2$ and $B=1T$, so that  $\varepsilon = 115.7 \mu eV$. We set $\Delta T_{dif}/T_{dif} = 10^{-1}$ as the upper bound for the relative shot-noise error. With a 10$\%$ temperature difference ($\beta_a/\beta_b = 1.1$), the operational temperature range is between 0.5 and 2.5 K. Therefore, for $T_a$=1K and $T_b$=1.1K, this method could achieve a resolution of 10 mK.

The lateral resolution of ESR-STM  thermal-gradient determination is determined by the smallest distance at which target atoms $a$ and $b$ can be placed. This distance is constrained by the capability of the ESR-STM active sensor to resolve three different peaks, associated to the states $GG$, $XG$ and $GX$. This can only happen if both the frequency shifts $\delta f_{a,b}= g_s\mu_B B_{a,b}$, associated to the stray fields created by atoms $a$ and $b$, $B_a$, $B_b$, as well as their difference, $\delta f_a-\delta f_b$ are larger than the peak linewidth $\delta f$. Linewidths as small as 3.6MHz  in  ESR-STM experiments have been reported \cite{baumann15}. 
For a sensor with $g=2$, $S=1/2$, the shift of the line is 28MHz per mT. In turn, the magnetic field created by a magnetic moment of 1$\mu_B$ at 1nm is $\approx 1.9$ mT. 
Thus, if we assume that both atoms $a$ and $b$ and the sensor atom have a magnetic moment of 1$\mu_B$, and their separation to the sensor atom is  $d + x$ and $d - x$, so that they are separated by a distance $2d$, and we 
take $d=1$ nm and $x=0.3$ nm, we find $2 |\delta f_a-\delta f_b|\simeq 64$ MHz, sufficient to resolve both peaks given a resonance width of 3.6 MHz \cite{baumann15}. Therefore, the lateral resolution of ESR-STM thermal gradients is in the range of 2 nm, dramatically outperforming scanning thermal magnetometry\cite{zhang20}, for which lateral resolutions are in the range of 100 nm. Combined with the 10 mK temperature difference relative error, ESR-STM offers a potential gradient resolution of $\frac{dT}{dx} \simeq 5$ mK/nm.

In conclusion, we have analyzed the working principles and the resolution limits of the ESR-STM quantum sensing technique to carry out direct temperature and temperature gradient measurements. We have emphasized that, unlike most temperature determination methods, ESR-STM provides a direct measurement of temperature, relying directly on the Boltzmann formula that relates temperature, energy and average occupation of quantum state. We have quantified the limits that both shot-noise and back-action set to the resolution and operational range of the methods, and we find that they dramatically outperform alternative techniques. Our theory relies on the lateral-sensing ESR-STM scheme that requires the resonant spin to be placed on an MgO surface, the standard situation so far \cite{baumann15,yang17,willke18,natterer19,seifert20,van21,farinacci22,kovarik22,zhang22,kot23,wang23}. However, the implementation of ESR-STM with the resonating spin placed in the STM tip, recently reported\cite{esat2024}, holds the promise of extending the application of the technique to any conducting surface, and therefore, opens the door for the implementation of direct temperature measurements with atomic-scale resolution, and may advance our understanding of many physical phenomena, including radiative heat transfer at the nanoscale\cite{cuevas18}.

\section*{Acknowledgements}

J.F.-R., Y. D. C.  
acknowledge financial support from 
%1
 FCT (Grant No. PTDC/FIS-MAC/2045/2021),
 %2
 SNF Sinergia (Grant Pimag),
 %3
 the European Union (Grant FUNLAYERS
- 101079184).
J.F.-R. acknowledges funding from 
% 4
Generalitat Valenciana (Prometeo2021/017
and MFA/2022/045)
%5
and
MICIN-Spain (Grants No. PID2019-109539GB-C41 and PRTR-C1y.I1)

\bibliography{bib}

\end{document}

\title{Theory of atomic-scale direct thermometry using ESR-STM: Supplemental Material}

\author{Y. del Castillo$^{1,2}$, J. Fern\'{a}ndez-Rossier$^1$\footnote{On permanent leave from Departamento de F\'{i}sica Aplicada, Universidad de Alicante, 03690 San Vicente del Raspeig, Spain}$^,$\footnote{joaquin.fernandez-rossier@inl.int} }

\affiliation{$^1$International Iberian Nanotechnology Laboratory (INL), Av. Mestre Jos\'{e} Veiga, 4715-330 Braga, Portugal }
\affiliation{$^2$Centro de F\'{i}sica das Universidades do Minho e do Porto, Universidade do Minho, Campus de Gualtar, 4710-057 Braga, Portugal }

\date{\today}
\maketitle
\appendix

\section{Estimate of resolution}
Here we
derive the relative error of the temperature measurement method using ESR-STM to determine both the ratio between the peaks and the splitting in the two-level system (TLS) (equation (9) and (10) of the main text). We start with the relation between temperature, ratio, and energy given by $k_B T = -\frac{\varepsilon}{ln (r)}$. We now apply the straightforward formula of error propagation with the two uncertainty sources, $r$ and $\varepsilon$

\begin{equation}
    \Delta k_B T= \left| \frac{\partial k_BT}{\partial \varepsilon} \right|\Delta \varepsilon
    +\left| \frac{\partial k_BT}{\partial r} \right| \Delta r
\end{equation}
then, we can trivially obtain
\begin{equation}
        \Delta k_B T = \left | \frac{1}{ln (r)}  \right | \Delta \varepsilon+ \left | \frac{\varepsilon }{r ln^2 (r)} \right |\Delta r .
\label{eq:error1}
\end{equation}
The uncertainty in the peak height ratio relates to the relative resolution in the current variations that can be registered. We showed in a previous work \cite{del23} that:
\begin{equation}
    \Delta r=\frac{\Delta I}{I_0}(1+r)
\label{eq:ratio_res}
\end{equation}
where $I_0$ is the base current and $\Delta I$ is the minimal current variation that can be registered, that is given by shot noise. In a given time interval $\Delta t$, an average of $N$ electrons goes through the STM-surface tunnel junction. Assuming a 
Poissonian distribution, the variance is given by  $\sqrt N$.

Therefore $\Delta I=\frac{e\sqrt{N}}{\delta t}= \sqrt{\frac{e I}{\Delta t}}$. We substitute in equation (\ref{eq:ratio_res}) and we have $\Delta r = (1+r)/ \sqrt{N}$ and combined with equation (\ref{eq:error1}), we obtain
\begin{equation}
        \Delta k_B T = \frac{\Delta \varepsilon}{|ln (r)|}  +\frac{\varepsilon (1+r)}{r \sqrt{N} |ln^2 (r)|}.
\end{equation}

After some algebra, we arrive at the following expression for the relative error:
\begin{equation}
       \frac{\Delta T}{T} =  \frac{\Delta \varepsilon}{\varepsilon} +\frac{1}{\sqrt{N}} \frac{1+r}{r   |\ln{r}|}.
\label{eq:resolution1}
\end{equation}
We can rewrite this equation in terms of inverse temperature and two-level splitting  using $r=e^{-\beta \varepsilon }$:

\begin{equation}
       \frac{\Delta T}{ T} = \frac{\Delta \varepsilon}{\varepsilon} +
      \frac{1}{\sqrt{N}}\frac{1+e^{\beta\varepsilon }}{\beta \varepsilon}.
\label{eq:resolution}
\end{equation}

\section{Role of back action}

We now discuss the role of back-action of the sensor spin on the occupation of the probe spins.  Even if flip-flop interactions are blocked, on account of the small ratio between their magnitude and the energy difference of the states $GX$ and $XG$, the Ising interaction $j S_z(s) S_z(p)$ is active, where $s$ and  $p$ stands for sensor and probe respectively. A lower bound for $j$ is the dipolar interaction:
\begin{equation}
    j=\frac{\mu_0 \mu_B^2 S_s S_p}{4\pi d^3}
\end{equation}

Hence, the effective Hamiltonian for the probe spin would be:
\begin{equation}
    H_{\rm probe}= g_p\mu_B B_{eff} S_z(p)  
\end{equation}
where 
\begin{equation}
 B_{eff}= B+ \frac{j S_z(s)}{g_p\mu_B}
 \end{equation}
in analogy with the effective Hamiltonian for the sensor spin.  Now, the main difference between these two situations is the different dynamics of the sensor and probe spins. The sensor spin is being driven by the AC perturbation, and it is therefore fluctuating in a time scale given by $\Omega^{-1}$.  In contrast, as long as the flip-flop interactions are blocked, the probe spin is fluctuating in a much slower time scale, given by $T_1$.  

The average magnetization of the driven sensor spin is given by\cite{delgado17}:
\begin{equation}
    \langle S_z(s)\rangle= \langle S_z\rangle_0\left(1- \frac{T_1 T_2 \Omega^2}{1+T_2^2\delta^2+T_1 T_2 \Omega^2}\right) 
\end{equation}
and $T_1, T_2, \Omega$ are the spin relaxation time, spin decoherence time and Rabi driving force of the sensor spin.
Thus, the average effective field acting on the probe is given by
\begin{equation}
\langle  B_{eff}\rangle = B+  \frac{j \langle S_z(s)\rangle}{g_p\mu_B}
 \end{equation}

At resonance, and in the limit $T_1 T_2 \Omega^2>>1$, we have  $\langle S_z(s)\rangle \rightarrow 0$.  However, given that the resonance value $\delta=0$ can only be achieved a fraction of the instances, on account of the fact that $\delta$ depends on the state in which the probe spin is, the effective field of the probe spin is {\em definitely different} from the external field by an amount {\em smaller}
than $\frac{j S_z(s)}{g_p\mu_B}$.  We can therefore provide the following bound to $\varepsilon$
\begin{equation}
    \Delta \varepsilon|_{\rm back action}\leq \frac{j S_z(s)}{g_p\mu_B}
\end{equation}
taking the most pessimistic assumption that the average spin of sensor spin is maximal. 
The relative error is given by the ratio of the external field and the stray field:
\begin{equation}
\frac{ \Delta \varepsilon|_{\rm back action}}{\varepsilon} \leq \frac{B_{\rm stray}}{B} \simeq 10^{-3}
\end{equation}
Therefore, the back-action error is clearly larger than the shot-noise limit when it comes to determining $\Delta \varepsilon$. 

In principle, back-action error does not depend on the measurement time. However, it is apparent that the detection of much smaller stray fields could be achieved if integration time is increased.  Another strategy would be to increase the external field, that would require to operate at larger frequencies.

\section{Engineered structures: non-interaction approximation}
The probabilities of changing the state of $n$ uncorrelated spins are given by the Boltzmann distribution using the combinatorial formula for ${\cal N}$ spins with each possible state where $n$ are excited:
\begin{equation}
    P_n(T)= \frac{\frac{{\cal N}!}{n! ({\cal N}-n)!} e^{-\beta \varepsilon_n}}{\sum_{k=0}^{{\cal N}} \frac{{\cal N}!}{k! ({\cal N}-k)!}e^{-\beta \varepsilon_k}}
\label{eq:prob_N}
\end{equation}
where $\varepsilon_n$ is the energy needed to flip $n$ spins simultaneously. We assume that the energy would be directly proportional to the energy of exciting a single spin, $\varepsilon_n=n \varepsilon_1$. 
\begin{equation}
\frac{\Delta T}{T} = \frac{\Delta \varepsilon}{\varepsilon} + \frac{1}{\sqrt{N}} \frac{r+1}{r |\ln{(r/{\cal N})}|} = \frac{\Delta \varepsilon}{\varepsilon} + \frac{1}{\sqrt{ N}}{\cal F}_{{\cal N}}(r)
\end{equation}
Similarly to equation \ref{eq:resolution}, we express the result in terms of $\beta \varepsilon$:
\begin{equation}
\frac{\Delta T}{T} = \frac{\Delta \varepsilon}{\varepsilon} + \frac{1}{\sqrt{ N}} \frac{{\cal N} + e^{\beta \varepsilon}}{ {\cal N} \beta \varepsilon} = \frac{\Delta \varepsilon}{\varepsilon} + \frac{1}{\sqrt{ N}}{\cal F}_{{\cal N}}(\beta \varepsilon)
\end{equation}
Moreover, we notice that for a high number of atoms surrounding the sensor equidistantly, the combinatorial formula may not work since correlations between neighboring spin would decrease the degeneracy for $n>1$, giving a higher probability to the first excited state, improving the precision at lower temperatures as long as the energy splitting due to neighboring interactions is distinguishable.

\section{Thermal gradient resolution}
In this section, we derive the relative temperature difference error (equation (22) in the main text). The temperature difference is given in terms of $\varepsilon$ and $R_a, R_b$:
\begin{equation}
    T_{dif} = -\frac{\varepsilon}{k_B} \bigg ( \frac{1}{ln R_a } - \frac{1}{ln R_b} \bigg )
\end{equation}
Then, using the error propagation formula, we get:
\begin{equation}
    \Delta T_{dif} = \left | \frac{\partial T_{dif}}{\partial \varepsilon } \right | \Delta \varepsilon +\left | \frac{\partial T_{dif}}{\partial R_a} \right | \Delta R_a + \left | \frac{\partial T_{dif}}{\partial R_b} \right | \Delta R_b
\end{equation}
Trivially, the term associated with the energy of the TLS gives:
\begin{equation}
    \left | \frac{\partial T_{dif}}{\partial \varepsilon } \right | \Delta \varepsilon = \frac{\Delta \varepsilon}{k_B} \left |  \frac{1}{ln{R_a} } - \frac{1}{ln{R_b}}  \right |
\end{equation}
Analogous to Appendix A, the terms related to the determination of peak height are as follows:
\begin{equation}
\begin{split}
    \left | \frac{\partial T_{dif}}{\partial R_a} \right | \Delta R_a &= \left | \frac{\varepsilon }{R_a ln^2 (R_a)} \right |\Delta R_a\\
    \left | \frac{\partial T_{dif}}{\partial R_b} \right | \Delta R_b &= \left | \frac{\varepsilon }{R_b ln^2 (R_b)} \right |\Delta R_b\\
\end{split}
\end{equation}
using $\Delta r = (1+r)/ \sqrt{N}$ we get the following expression:
\begin{equation}
    \Delta T_{dif} = \frac{\Delta \varepsilon}{k_B} \left |  \frac{1}{ln{R_a} } - \frac{1}{ln{R_b}}  \right | + \left | \frac{\varepsilon }{k_B R_a ln^2 (R_a)} \right |\Delta R_a + \left | \frac{\varepsilon }{k_B R_b ln^2 (R_b)} \right |\Delta R_b
\end{equation}
and after some rearrangement of the equation, we find
\begin{equation}
    \Delta T_{dif}= \frac{\Delta \varepsilon}{ k_B\varepsilon} \left |  \frac{\beta_b - \beta_a}{\beta_a \beta_b }  \right | +  \frac{1}{\sqrt{N}\varepsilon k_B} \left ( \frac{1+R_a}{ R_a \beta_a^2 }  +  \frac{1+R_b}{R_b \beta_b^2 } \right ) 
\end{equation}
Now, we combine $\Delta T_{dif}$ with $T_{dif}$ in terms of $\beta_{a,b}$, $T_{dif} = \frac{1}{k_B} \left ( \frac{\beta_b-\beta_a}{\beta_a \beta_b } \right )$. The ratio between both expressions gives us the relative error for temperature differences for the gradient determination method:
\begin{equation}
    \frac{\Delta T_{dif}}{T_{dif}} = \frac{\Delta \varepsilon}{\varepsilon} + \frac{\beta_a \beta_b}{\sqrt{N} \varepsilon \left( \beta_b - \beta_a \right)} \left( \frac{1 + R_a}{R_a \beta_a^2} + \frac{1 + R_b}{R_b \beta_b^2} \right)
\end{equation}

Using $R_a = e^{-\beta_a \varepsilon}$ and $R_b = e^{-\beta_b \varepsilon}$, analogously to the relative error from equations (9) and (10) in the main text, we get a physical form of the expression:
\begin{equation}
    \frac{\Delta T_{dif}}{T_{dif}} = \frac{\Delta \varepsilon}{\varepsilon} + \frac{\beta_a \beta_b}{\sqrt{N} \varepsilon (\beta_b - \beta_a)} \left( \frac{e^{\beta_a \varepsilon} + 1}{\beta_a^2} + \frac{e^{\beta_b \varepsilon} + 1}{\beta_b^2} \right)
\end{equation}
 and the operational form in terms of the ratios, $R_a,R_b$:
\begin{equation}
    \frac{\Delta T_{dif}}{T_{dif}} = \frac{\Delta \varepsilon}{\varepsilon} + \frac{ln (R_a)ln (R_b) }{\sqrt{N}  ln \left (R_b/ R_a\right )} \left( \frac{1 + R_a}{R_a ln^2 (R_a) } + \frac{1 + R_b}{R_b ln^2 (R_b) } \right).
\end{equation}

where the second part relates to the determination of the ratio and it only depends on the ratios measured and the shot noise, $\frac{1}{\sqrt{N}} {\cal F}_{dif}(R_a,R_b)$.

\bibliography{bib}